\documentclass[preprint,showpacs,preprintnumbers,amsmath,amssymb,prb,aps]{revtex4-1}
\usepackage{epsfig}
\usepackage{dcolumn}

\begin{document}

\title{The role of ionic sizes in inducing the cubic to tetragonal distortion in AV$_{2}$O$_{4}$ and ACr$_{2}$O$_{4}$ (A=Zn, Mg and Cd) compounds}
\author{Sohan Lal} 
\altaffiliation{Electronic mail:goluthakur2007@gmail.com}
\author{Sudhir K. Pandey}
\affiliation{School of Engineering, Indian Institute of Technology Mandi, Kamand 175005, Himachal Pradesh, India}

\date{\today}

\begin{abstract}

   Cubic to tetragonal distortion in AV$_{2}$O$_{4}$ and ACr$_{2}$O$_{4}$ (A=Zn, Mg and Cd) compounds have been a contentious issue for last two decades. Different groups have proposed different mechanisms to understand such a distortion in these spinels, which are: (i) spin lattice coupling mechanism known as the spin driven Jahn-Teller (JT) effect, (ii) the strong relativistic spin-orbit coupling, a moderate JT distortion and weak V-V interactions and (iii) the JT effect. Now, in order to know the possible cause for such a distortion, we have avoided these complexities (various interactions among spin, electronic, orbital and lattice degrees of freedom) by carrying out spin unpolarized calculations. The calculated values of bulk moduli for ZnV$_{2}$O$_{4}$ (ZnCr$_{2}$O$_{4}$), MgV$_{2}$O$_{4}$ (MgCr$_{2}$O$_{4}$) and CdV$_{2}$O$_{4}$ (CdCr$_{2}$O$_{4}$) are found to be $\sim$289 ($\sim$254), $\sim$244 ($\sim$243) and $\sim$230 ($\sim$233) GPa, respectively which suggest that CdV$_{2}$O$_{4}$ (among vanadates) and CdCr$_{2}$O$_{4}$ (among chromates) are more compressible. For vanadates and chromates, the order of calculated values of lattice parameter $a$ are found to CdV$_{2}$O$_{4}$$>$MgV$_{2}$O$_{4}$$>$ZnV$_{2}$O$_{4}$ and CdCr$_{2}$O$_{4}$$>$MgCr$_{2}$O$_{4}$$>$ZnCr$_{2}$O$_{4}$, respectively and are consistent with the experimental results. The calculated values of cubic to tetragonal distortion (c/a), with c/a$<$1 for ZnV$_{2}$O$_{4}$ (ZnCr$_{2}$O$_{4}$), MgV$_{2}$O$_{4}$ (MgCr$_{2}$O$_{4}$) and CdV$_{2}$O$_{4}$ (CdCr$_{2}$O$_{4}$) are $\sim$0.996 ($\sim$0.997), $\sim$0.995 ($\sim$0.994) and $\sim$0.997 ($\sim$0.998), respectively. These values are in good agreement with the experimental data for ZnV$_{2}$O$_{4}$, MgV$_{2}$O$_{4}$, ZnCr$_{2}$O$_{4}$ and MgCr$_{2}$O$_{4}$ compounds. The present study clearly shows the role of ionic sizes in inducing the cubic to tetragonal distortion in these spinels. However, the discrepancies between the calculated and experimental data for CdV$_{2}$O$_{4}$ and CdCr$_{2}$O$_{4}$ are expected to improve by considering the above mentioned mechanisms. These mechanisms also appear to be responsible for deciding the other physical properties of these compounds.

\end{abstract}

\maketitle

\section{Introduction} 
  
    In the transition metal oxides, the $d$ level is fivefold degenerate. The degeneracy of $d$ level is split into the lower energy t$_{2g}$ level (with degenerate $d$$_{xy}$, $d$$_{xz}$ and $d$$_{yz}$ orbitals) and higher energy e$_{g}$ level (with degenerate $d$$_{x^{2}-y^{2}}$ and $d$$_{z^{2}}$ orbitals) by the crystal field splitting in an octahedral field. Normally, in some of the transition metal oxides, Jahn-Teller (JT) effect has been mainly attributed for a structural transition, which take place from high temperature cubic phase to low temperature phases of the compound.\cite{Goodenough} It is important to note that the JT distortion can lead both elongation and compression of octahedra depending on the number of $d$ electrons.\cite{Khomskii2014} Here, we discuss only the physics of d$^{2}$ and d$^{3}$ electron systems. In d$^{2}$ electron systems, two of the threefold degenerate t$_{2g}$ orbitals are occupied, whereas all three orbitals are occupied for d$^{3}$ electron systems. Now, for d$^{2}$ electron systems, the degeneracy of t$_{2g}$ level can be lifted by two ways. First one by JT mechanism and second one by both JT and spin-orbit mechanisms.\cite{Khomskii2014,Kugel} In these systems, tetragonal elongation of the local octahedron with, c/a$>$1 is more favorable from the viewpoint of JT effect, where both d electrons are occupied by lowest energy degenerate $d$$_{xz}$ and $d$$_{yz}$ orbitals.\cite{Khomskii2014} In such a situation, the ground state has unfrozen orbital angular momentum. Hence, a further splitting of lowest doublet is expected due to the spin-orbit interaction.\cite{Khomskii2014,Kugel} For d$^{3}$ electron systems, JT effect is inactive and hence the driving force for structural transition is something of different nature than d$^{2}$ electron systems.  
    
    Spinel compounds, AV$_{2}$O$_{4}$ and ACr$_{2}$O$_{4}$ (A=Zn, Mg and Cd) with the face-centered-cubic structure at room temperature are among the most extensively studied geometrically frustrated magnets.\cite{Dutton,Yamashita,Maitra,Lee,Onoda,Reehuis,Wheeler,Mamiya,Nishiguchi,Tsunetsugu,Tchernyshyov,Matteo,Khomskii,Pandey2011,Pandey2012, Lal2014,Lal2016,Giovannetti,Suzuki} 
Vanadium spinels (AV$_{2}$O$_{4}$) and chromium spinels (ACr$_{2}$O$_{4}$) crystallize in a face-centered cubic structure, where V and Cr ions are occupied at the octahedral sites, respectively. In vanadium and chromium spinels, a pyrocholre lattice is formed due to the corner sharing tetrahedral network of magnetically coupled V and Cr ions, respectively.\cite{Lee,Suzuki,Broholm1,Broholm2,Rovers,Shaked} Vanadium spinels with V$^{3+}$ (3$d$$^{2}$, $S$=1) ion is JT active, whereas chromium spinels with Cr$^{3+}$ (3$d$$^{3}$, $S$=3/2) ion is JT inactive as per expectation as discussed above. Hence for vanadates, JT effect is expected to be responsible for a cubic to tetragonal structural transition with c/a$>$1, which is contrary to the experimental result.\cite{Onoda,Reehuis,Wheeler} However, for chromium spinels, JT effect is not expected to play any role in inducing such a transition. Structural transition from cubic to tetragonal are common in these spinels, which has been a long issue from last two decades.\cite{Tchernyshyov,Khomskii,Tsunetsugu,Yamashita,Moessner} In vanadium spinels, some of the researcher have attributed the JT effect responsible for cubic to tetragonal structural transition.\cite{Tchernyshyov,Tsunetsugu} The experimentally reported values of structural transition temperature, $T$$_{S}$ (tetragonal distortion, c/a) for CdV$_{2}$O$_{4}$, MgV$_{2}$O$_{4}$ and ZnV$_{2}$O$_{4}$ compounds are $\sim$97 K (0.9877), $\sim$65 K (0.9941) and $\sim$50 K (0.9948), respectively.\cite{Mamiya,Nishiguchi,Reehuis,Onoda,Wheeler} In these vanadates, it is interesting that the magnetic transition occur at a temperature less than $T$$_{S}$.\cite{Mamiya,Nishiguchi,Reehuis,Onoda,Wheeler} However, in chromium spinels, both structural and magnetic transition take place at the same temperature. Hence, the magnetostructural transition temperature, $T$$_{S}$$\approx$$T$$_{N}$ (tetragonal distortion, c/a) for CdCr$_{2}$O$_{4}$, MgCr$_{2}$O$_{4}$ and ZnCr$_{2}$O$_{4}$ are $\sim$7.8 K (1.004), $\sim$12.5 K (0.9979) and $\sim$12.5 K (0.9981), respectively.\cite{Martin,Dutton,Kagomiya,Chung,Takagi} This makes these chromium spinels distinct from the vanadium spinels. Another interesting thing in these spinels is that the structural transition, with c/a$>$1 is reported experimentally only for CdCr$_{2}$O$_{4}$, whereas it is reported to be c/a$<$1 for remaining five compounds.
    
    In order to understand a long issue of cubic to tetragonal structural transition in these spinels, different groups have proposed different theories.\cite{Yamashita,Tsunetsugu,Tchernyshyov,Khomskii,Moessner} First, we start with the vanadium spinels, which is more controversial for such a transition. Based on the valence-bond-solid approach, Yamashita and Ueda proposed a spin-JT coupling mechanism that leads to a cubic to tetragonal structural transition for ZnV$_{2}$O$_{4}$ and MgV$_{2}$O$_{4}$ compounds.\cite{Yamashita} However, based on this approach, it is difficult to explain the magnetic transition at a temperature less than $T$$_{S}$. In the light of this model, Tsunetsugu and Motome have proposed a scenario to explain such a transition in AV$_{2}$O$_{4}$. In their scenario, this transition is found to be an orbital order accommodated by the JT distortion, with c/a$<$1.\cite{Tsunetsugu} However, their model is incompatible with the experimentally observed spatial symmetry {\it I}4$_{1}$/{\it amd} of the tetragonal phase as it breaks the mirror reflections in the planes (110) and (1$\bar{1}$0) and diamond glides {\it d} in the planes (100) and (010).\cite{Reehuis,Nishiguchi}  Tchernyshyov also offered a theoretical model based on the relativistic spin-orbit coupling, collective JT effect and spin frustration to explain the transition in the spinel vanadates. Their model is divided into two cases. In the first case JT coupling is considered as a dominating factor, which leads a cubic to tetragonal transition with c/a$>$1 as per expectation, as discussed above. Such a transition with c/a$>$1 is incompatible with the experimental result.\cite{Onoda,Reehuis,Wheeler} In the second case, Tchernyshyov has considered strong relativistic spin-orbit coupling, a moderate JT distortion and weak V-V interaction that yield a cubic to tetragonal transition with c/a$<$1, compatible with the experimental result.\cite{Onoda,Reehuis,Wheeler,Tchernyshyov} Khomskii {\it et al.} proposed a model, where cubic to tetragonal structural transition is observed due to the anti-JT effect caused by the broadening of the yz, xz bands leads to the orbitally driven Peierls state.\cite{Khomskii} For chromium spinels, Yamashita {\it et al.} and Tchernyshyov {\it et al.} have proposed a model to explain such a transition, which is based on the spin-driven JT effect. According to their model, the strong spin degeneracy due to the geometrical frustration is lifted by spin-driven JT effect (lattice distorts spontaneously and the system switches from paramagnetic spin liquid to antiferromagnetic order state) and leads to a cubic to tetragonal structural transition.\cite{Yamashita,Moessner}

   The crystal structure of a material is made up of regular arrangements of atoms or ions with well defined radii (which is a measure of the size of atoms or ions). It is well known that many of the physical properties of a material depends on the crystal structure and hence is one of the most important aspects of the solid state physics.\cite{Ashcroft} Now, in order to understand the various physical properties of the compound, it is important to include the effect of atomic or ionic sizes. In model calculations, inclusion of the effect of atomic or ionic sizes is not a straightforward job. However, in {\it ab initio} electronic structure calculations, it is included inherently. Here, it is important to note that the role of ionic sizes for deciding the cubic to tetragonal structural transition is expected for above mentioned spinels. However, none of the groups have discussed this aspect in these spinels. 
   
   From above discussion, it is clear that the cubic to tetragonal distortion is yet controversial for these spinels. Here, we have tried to understand such a distortion by considering only the effect of ionic sizes, where we have ignored  all the complexities (as discussed above) by performing spin unpolarized calculations for these compounds. The calculated value of bulk modulus is found to be smallest for CdV$_{2}$O$_{4}$ (among AV$_{2}$O$_{4}$) and CdCr$_{2}$O$_{4}$ (among ACr$_{2}$O$_{4}$) which suggest that the CdV$_{2}$O$_{4}$ and CdCr$_{2}$O$_{4}$ are more compressible as compared to other vanadium and chromium spinels, respectively. Local density approximation (LDA) exchange correlation functional underestimates the calculated values of equilibrium lattice parameter $a$ in the range of 2.4-3.3\% as compared to the experimental results for these compounds. For vanadates, the calculated values of c/a, with c/a$<$1 for ZnV$_{2}$O$_{4}$ and MgV$_{2}$O$_{4}$ are close to the experimental data. However, for CdV$_{2}$O$_{4}$, it is deviated by $\sim$0.9\% from the experimental result. For chromium spinels, the deviation of c/a from the experimental data are found to be $\sim$0.1\%, $\sim$0.4\% and $\sim$0.6\% for ZnCr$_{2}$O$_{4}$, MgCr$_{2}$O$_{4}$ and CdCr$_{2}$O$_{4}$, respectively. The present study clearly shows that the main cause for cubic to tetragonal distortion in these spinels are the effect of ionic sizes. However, above mentioned mechanisms may be responsible for the other physical properties of these spinels.

\section{Computational Detail}       
    
    The spin unpolarized (SUP) calculations of six spinel compounds, AV$_{2}$O$_{4}$ and ACr$_{2}$O$_{4}$ (A=Zn, Mg and Cd) are carried out by using the full-potential linearized-augmented plane-wave (FP-LAPW) method as implemented in elk code.\cite{elk} The calculations for every compounds are performed in the face centered cubic phase. The lattice parameters and atomic positions for these compounds are taken from the literature.\cite{Onoda,Reehuis,Martin,Chung,Dutton,Wheeler} LDA, Perdew -Wang/Ceperley -Alder exchange correlation functional has been used in these calculations.\cite{Perdew} The muffin-tin sphere radii (in Bohr) used in the calculations for every compounds are given in the Table I. We have used the 8x8x8 k-point grid size. In all calculations, the basis set cut off muffin-tin radius times maximum $\lvert$G+k$\lvert$ (rgkmax) and maximum length of $\lvert$G$\lvert$ for expending the interstitial density and potential (gmaxvr) are set to be 8.0 and 14.0, respectively. These values are good for obtaining the fine parabolic curves of energy versus volume. Convergence target of total energy has been set below 10$^{-4}$ Hartree/cell.  
   
    Now, in order to know the equilibrium lattice parameters, we have done the full structure optimization for every compounds. Atomic positions corresponding to the relaxed structure of these compounds are fixed during the calculation of lattice parameters. The equilibrium lattice parameter is calculated by fitting the total energy formula unit versus unit cell volume data using the universal equation of state.\cite{Vinett} The universal equation of state is defined as,
     
     $P$ = [3$B$$_{0}$(1 - $\chi$)/$\chi$$^{2}$]e$^{3/2(B'_{0}-1)(1-\chi)}$, $P$ = -($\partial$$E$/$\partial$$V$)   
    where $P$, $E$, $V$, $B$$_{0}$ and $B$$_{0}$$^{'}$ are the pressure, energy, volume, bulk modulus and pressure derivative of bulk modulus, respectively and $\chi$ = ($V$/$V_{0}$)$^{1/3}$.

\section{Result and Discussion} 

    First of all, we discuss the crystal structure of AV$_{2}$O$_{4}$ and ACr$_{2}$O$_{4}$ (A=Zn, Mg and Cd) spinel compounds. Except MgV$_{2}$O$_{4}$, all five compounds crystallize in face centered cubic spinel structure with the space group {\it Fd$\bar{3}$m}. In these five compounds, (Zn, Mg, Cd) and (V, Cr) atoms are found at the Wyckoff positions 8$a$ (0.125,0.125,0.125) and 16$d$ (0.5,0.5,0.5), respectively. The O atom is located at the Wyckoff position 32$e$ (x,x,x), where the values of x for five different compounds are shown in the Table II. However, MgV$_{2}$O$_{4}$ crystallizes in the face centered cubic structure characterized by space group {\it F$\bar{4}$3m}. In this compound, Mg atom is located at the Wyckoff positions 4$a$ (0,0,0) and 4$c$ (0.25,0.25,0.25). However, both V and O (O1 and O2) atoms are located at the Wyckoff position 16$e$ (x,x,x), where the values of x are also shown in the Table II.       
     
    Total energy difference between the volume dependent energies and energy corresponding to the equilibrium volume [$\Delta$$E$=$E$(V)-$E$(V$_{\rm eq}$)] per formula unit versus unit cell volume plots for all compounds in the cubic phase obtained from the SUP LDA calculations are shown in the Fig. 1(a-f). In these compounds, each curve shows almost a parabolic behavior and the volume corresponding to the minimum energy gives the equilibrium volume. In order to determine the equilibrium volumes for these compounds, we have fitted the total energy-volume data by the using universal equation of state.\cite{Vinett} The equilibrium volumes for ZnV$_{2}$O$_{4}$, MgV$_{2}$O$_{4}$ and CdV$_{2}$O$_{4}$ compounds are $\sim$3617.4, $\sim$3662.6 and $\sim$4037.3 bohr$^{3}$, respectively. Similarly, for ZnCr$_{2}$O$_{4}$, MgCr$_{2}$O$_{4}$ and CdCr$_{2}$O$_{4}$, its values are $\sim$3558.8, $\sim$3601 and $\sim$3980.3 bohr$^{3}$, respectively. The equilibrium values of lattice parameter ($a$) are obtained from the equilibrium volumes for these compounds. The values of $a$ for ZnV$_{2}$O$_{4}$, MgV$_{2}$O$_{4}$ and CdV$_{2}$O$_{4}$ are $\sim$8.125 {\AA}, $\sim$8.157 {\AA} and $\sim$8.426 {\AA}, respectively. Similarly, for ZnCr$_{2}$O$_{4}$, MgCr$_{2}$O$_{4}$ and CdCr$_{2}$O$_{4}$ compounds, its values are $\sim$8.079 {\AA}, $\sim$8.111 {\AA} and $\sim$8.386 {\AA}, respectively. Now, we compare the calculated values of $a$ with experimentally observed values of $a$ for these spinel compounds. The calculated and experimentally reported (shown in bracket) values of $a$ for these spinels are shown in the Table II. It is clear from the table that for vanadium spinels, the order of calculated values of $a$ is similar to the experimental one, which is CdV$_{2}$O$_{4}$$>$MgV$_{2}$O$_{4}$$>$ZnV$_{2}$O$_{4}$. Similarly, for chromium spinels, the calculated values of $a$ for CdCr$_{2}$O$_{4}$$>$MgCr$_{2}$O$_{4}$$>$ZnCr$_{2}$O$_{4}$, which are consistent with the experimentally reported order. The experimentally observed values of $a$ for ZnV$_{2}$O$_{4}$, MgV$_{2}$O$_{4}$ and CdV$_{2}$O$_{4}$ are 8.4028 {\AA}, 8.42022 {\AA} and 8.691 {\AA}, respectively, which are $\sim$3.3{\%}, $\sim$3.1{\%} and $\sim$3.0{\%} greater than the calculated one. Similarly, for ZnCr$_{2}$O$_{4}$, MgCr$_{2}$O$_{4}$ and CdCr$_{2}$O$_{4}$ compound, its values are 8.320721 {\AA}, 8.3329 {\AA} and 8.59093 {\AA}, respectively, which are $\sim$2.9{\%}, $\sim$2.6{\%} and $\sim$2.4{\%} larger than the calculated results. Here, in the present study, the large underestimation of the calculated values of $a$ as compared to experimental results for these spinels are due to the following reasons: (i) it is well known that the LDA method itself underestimates the lattice parameters\cite{Haas} and (ii) in present work, we have performed the SUP LDA calculations, which further underestimate the values of lattice parameter.  
     
    Now, we discuss the calculated values of bulk moduli for the above mentioned compounds, which are shown in the Table II. It is evident from the table that among vanadium spinels, the bulk modulus is largest for ZnV$_{2}$O$_{4}$ and smallest for CdV$_{2}$O$_{4}$. Its values for ZnV$_{2}$O$_{4}$, MgV$_{2}$O$_{4}$ and CdV$_{2}$O$_{4}$ compounds are $\sim$289 GPa, $\sim$244 GPa and $\sim$230 GPa, respectively. The calculated value of bulk modulus for CdV$_{2}$O$_{4}$ in the present study is $\sim$60-80 GPa more than that calculated by two different groups, where they have used the GGA and hybrid functionals.\cite{Pandey2012,Canosa} However, on the basis of our knowledge, the values of bulk moduli for both ZnV$_{2}$O$_{4}$ and MgV$_{2}$O$_{4}$ compounds are not reported experimentally and theoretically. Due to which, we can not compare our results for both spinels. Similarly, among chromium spinels, the calculated value of bulk modulus is largest for ZnCr$_{2}$O$_{4}$ and smallest for CdCr$_{2}$O$_{4}$. The values of bulk moduli for ZnCr$_{2}$O$_{4}$, MgCr$_{2}$O$_{4}$ and CdCr$_{2}$O$_{4}$ compounds are $\sim$254 GPa, $\sim$243 GPa and $\sim$233 GPa, respectively. Here, we compare the present calculated values of bulk moduli with the bulk moduli predicted by different groups. Theoretically, Catti $et$ $al$. have predicted the bulk moduli 215 GPa and 197.3 GPa for ZnCr$_{2}$O$_{4}$ and MgCr$_{2}$O$_{4}$, respectively, where they have used the Hartree-Fock approach.\cite{Catti} However, the experimentally reported values of bulk moduli for ZnCr$_{2}$O$_{4}$ and MgCr$_{2}$O$_{4}$ are 183.1 GPa and 189 GPa, respectively.\cite{Yong} The order of bulk moduli for both compounds in the present study is similar to that predicted by Catti $et$ $al$, but different from the experimental results. The calculated values of bulk moduli for ZnCr$_{2}$O$_{4}$ and MgCr$_{2}$O$_{4}$ in this work are about $\sim$38{\%} ($\sim$18{\%}) and $\sim$28{\%} ($\sim$23{\%}) larger than the experimental (theoretical, predicted by Catti $et$ $al$.) results, respectively. The various reason for such an overestimation of the bulk moduli for these spinels in the present study is discussed below: (i) in general, LDA exchange-correlation functional has been found to overestimates the values of bulk moduli,\cite{Csonka} (ii) bulk modulus is also quite sensitive for various parameters used in the calculations.\cite{Lejaeghere} In the present study, we have performed only SUP calculations. However, spin polarized calculations are expected to improve the values of bulk moduli and (iii) different experimental techniques give the different values of bulk moduli. For example, Reichmann $et$ $al$. have observed the bulk modulus 185.7 GPa for FeFe$_{2}$O$_{4}$ using gigahertz ultrasonic interferometry and single crystal X-ray diffraction techniques.\cite{Reichmann} However, Haavik $et$ $al$. have reported the bulk modulus 217 GPa for FeFe$_{2}$O$_{4}$ using X-ray diffraction, where they have fitted the pressure-volume data using a third-order Birch-Murnaghan equation of state.\cite{Haavik}
    
   The bulk modulus of the crystal is related to the strength of its constituent bonds. For AV$_{2}$O$_{4}$ compounds, the bond lengths of the constituents depend on the ionic sizes of the A site (Zn$^{2+}$, Mg$^{2+}$ and Cd$^{2+}$). Nishiguchi $et$ $al$. have shown experimentally that the V-O bond length for these vanadates do not depend on the A site. However, they have observed that both V-V distance and the V-O-V angle decreases as the A site changes from Cd$^{2+}$, Mg$^{2+}$ and Zn$^{2+}$.\cite{Nishiguchi} Hence, if we ascribe such a dependency of these compositions to the difference in the ionic radii at A site, then the bulk modulus is expected to be less for CdV$_{2}$O$_{4}$ as compared to MgV$_{2}$O$_{4}$ and ZnV$_{2}$O$_{4}$ compounds. This is because of the large ionic radius of Cd$^{2+}$ (means easy to compress) as compared to Zn$^{2+}$ and Mg$^{2+}$. Hence, among vanadium spinels, the calculated value of bulk modulus is smallest for CdV$_{2}$O$_{4}$, whereas among chromium spinels, it is smallest for CdCr$_{2}$O$_{4}$ as compared to other compounds.

   The calculated values of x (represent the $x$, $y$ and $z$ coordinates of the atom) for O and V (only for MgV$_{2}$O$_{4}$) atoms corresponding to the calculated equilibrium lattice parameter $a$ of the above mentioned compounds are shown in the Table II. In the Table II, experimentally reported values of x for O and V (only for MgV$_{2}$O$_{4}$) atoms are also shown in the bracket. It is evident from the table that for vanadium spinels, the calculated values of x for O atom are $\sim$0.257 and $\sim$0.267 for ZnV$_{2}$O$_{4}$ and CdV$_{2}$O$_{4}$ compounds, respectively. However, for MgV$_{2}$O$_{4}$, the values of x for O1, O2 and V atoms are $\sim$0.387, $\sim$0.871 and $\sim$0.635, respectively. For vanadium spinels, the calculated values of x for O atom are deviated by $\sim$1.3\% and $\sim$0.07\% from experimental one for ZnV$_{2}$O$_{4}$ and CdV$_{2}$O$_{4}$ compounds, respectively. However, for MgV$_{2}$O$_{4}$, these are deviated from the experimental results by $\sim$0.2\%, $\sim$0.5\% and $\sim$1.6\% for O1, O2 and V atoms, respectively. Similarly, for chromium spinels, the calculated values of x for O atom are $\sim$0.258, $\sim$0.259 and $\sim$0.269 for ZnCr$_{2}$O$_{4}$, MgCr$_{2}$O$_{4}$ and CdCr$_{2}$O$_{4}$ compounds, respectively. Hence, the calculated values of x for O atom are deviated by $\sim$1.3\%, $\sim$0.8\% and $\sim$0.3\% from the experimental results for ZnCr$_{2}$O$_{4}$, MgCr$_{2}$O$_{4}$ and CdCr$_{2}$O$_{4}$ compounds, respectively. 
    
    In order to study the cubic to tetragonal distortion in the above mentioned compounds, we have fixed the equilibrium values of lattice parameter $a$=$b$ and have varied the parameter $c$. The plots of the total energy difference between the volume dependent energies and energy corresponding to the equilibrium volume [$\Delta$$E$=$E$(V)-$E$(V$_{\rm eq}$)] per formula unit versus percentage change in the calculated values of c/a (denoted by c/a\%) are shown in the Fig. 2(a-f). It is clear from the figure that every plot corresponding to the every compound shows almost a parabolic behaviour. The c/a\% corresponding to the minimum energy provides the cubic to tetragonal distortion in these compounds, where c/a\%=0 means no distortion. It is also clear from the figure that for all six compounds, the cubic to tetragonal distortion are finite, with c/a$<$1. Except for CdCr$_{2}$O$_{4}$, such a distortion with c/a$<$1 are compatible with experimental results for remaining five compounds. The experimentally observed (shown in the bracket) and calculated (corresponding to the minimum energy) values of c/a for these spinels are also shown in the Table II. It is evident from the table that for vanadium spinels, the calculated values of c/a for ZnV$_{2}$O$_{4}$, MgV$_{2}$O$_{4}$ and CdV$_{2}$O$_{4}$ are $\sim$0.996, $\sim$0.995 and $\sim$0.997, respectively. However, the experimentally observed values of c/a for ZnV$_{2}$O$_{4}$, MgV$_{2}$O$_{4}$ and CdV$_{2}$O$_{4}$ are 0.9948, 0.9941 and 0.9877, respectively, which suggest that the calculated values of c/a are in good agreement with the experimental results for both ZnV$_{2}$O$_{4}$ and MgV$_{2}$O$_{4}$ compounds. However, for CdV$_{2}$O$_{4}$, it is $\sim$0.9\% deviated from the experimental data. Similarly for chromium spinels, the calculated values of c/a for ZnCr$_{2}$O$_{4}$, MgCr$_{2}$O$_{4}$ and CdCr$_{2}$O$_{4}$ are $\sim$0.997, $\sim$0.994 and $\sim$0.998, respectively. The experimentally reported values of c/a for ZnCr$_{2}$O$_{4}$, MgCr$_{2}$O$_{4}$ and CdCr$_{2}$O$_{4}$ are 0.9981, 0.9979 and 1.004, respectively. The calculated value of c/a for ZnCr$_{2}$O$_{4}$ is in good agreement with the experimental one. However, for MgCr$_{2}$O$_{4}$ and CdCr$_{2}$O$_{4}$, its values are $\sim$0.4\% and $\sim$0.6\% deviated from the experimental results, respectively. It is also clear from the figure that the small change in the energy occur as structure changes from cubic to tetragonal for all six compounds. However, the various proposed mechanisms (as discussed in introduction) may be responsible to enhance the change in energy between cubic and tetragonal structures for these spinels.
    
    Now, in order to see the difference between cubic and tetragonal structures of these spinels, we have plotted the partial density of states (PDOS) of V and Cr atoms for these compounds. PDOS of V atom below Fermi level (zero energy) for both cubic and tetragonal structures of ZnV$_{2}$O$_{4}$, MgV$_{2}$O$_{4}$ and CdV$_{2}$O$_{4}$ are shown in the Fig. 3(a-c). Similarly, PDOS of Cr atom below Fermi level for both cubic and tetragonal structures of ZnCr$_{2}$O$_{4}$, MgCr$_{2}$O$_{4}$ and CdCr$_{2}$O$_{4}$ are shown in the Fig. 4(a-c). PDOS of V and Cr atoms for both structures are obtained in the SUP calculations by using the equilibrium values of lattice parameters and atomic coordinates of these compounds. The V and Cr 3$d$ states are mainly contributed to the PDOS of V (for vanadium spinels) and Cr (for chromium spinels) atoms, respectively. It is clear from the Fig. 3 that below Fermi level, PDOS of V atom for cubic and tetragonal structures of vanadium spinels are different. The contribution of V 3$d$ states to the PDOS of V atom for tetragonal structure are less than the cubic structure of these vanadates for whole energy range (shown in figure) below Fermi level. Similar behavior is also observed in the chromium spinels as shown in Fig. 4, where the contribution of Cr 3$d$ states to the PDOS of Cr atom below the Fermi level (for whole energy range) for tetragonal structure are also less than the cubic structure. Above discussion clearly shows that the contribution of $d$ states to the PDOS of V and Cr atoms for these spinels decreases as structure changes from cubic to tetragonal. From above discussion, it is also clear that the ionic sizes appear to be responsible for the cubic to tetragonal distortion for all six compounds. Even having the importance of ionic sizes for such a distortion, none of the groups have discussed this aspect. Interesting thing in the present study is that the calculated values of c/a (with c/a$<$1) for ZnV$_{2}$O$_{4}$, MgV$_{2}$O$_{4}$, ZnCr$_{2}$O$_{4}$ and MgCr$_{2}$O$_{4}$ compounds are good matching with the experimental results even by neglecting the complexities of various parameters as mentioned below: (i) spin lattice coupling mechanism known as the spin driven JT effect,\cite{Yamashita,Moessner} (ii) the strong relativistic spin-orbit coupling, a moderate JT distortion and weak V-V interactions\cite{Tchernyshyov} and (iii) the JT effect.\cite{Tsunetsugu} These mechanisms are discussed in more details in the introduction. The large deviation of calculated values of c/a, with c/a$<$1 for both CdV$_{2}$O$_{4}$ and CdCr$_{2}$O$_{4}$ from the experimental results, with c/a$<$1 for CdV$_{2}$O$_{4}$ and c/a$>$1 for CdCr$_{2}$O$_{4}$ is expected to improve by these mechanisms. These mechanisms may also be responsible for deciding the other physical properties like, space group and coincidence of structural and magnetic transitions etc. At last, we conclude that the present study clearly shows the importance of ionic sizes in inducing the cubic to tetragonal distortion in these spinels.

\section{Conclusions} 

   A long issue of cubic to tetragonal distortion in AV$_{2}$O$_{4}$ and ACr$_{2}$O$_{4}$ (A=Zn, Mg and Cd) compounds have been studied by different mechanisms proposed by various groups. These mechanisms are listed as: (i) spin lattice coupling mechanism known as the spin driven JT effect, (ii) the strong relativistic spin-orbit coupling, a moderate JT distortion and weak V-V interactions and (iii) the JT effect. In order to understand the possible cause behind such a distortion, we have only considered the effect of ionic sizes and have ignored above complexities by performing spin unpolarized {\it ab initio} electronic structure calculations. The order of calculated values of bulk moduli for vanadium and chromium spinels were found to ZnV$_{2}$O$_{4}$$>$MgV$_{2}$O$_{4}$$>$CdV$_{2}$O$_{4}$ and ZnCr$_{2}$O$_{4}$$>$MgCr$_{2}$O$_{4}$$>$CdCr$_{2}$O$_{4}$, respectively. The calculated values of lattice parameter $a$ for ZnV$_{2}$O$_{4}$ (ZnCr$_{2}$O$_{4}$), MgV$_{2}$O$_{4}$ (MgCr$_{2}$O$_{4}$) and CdV$_{2}$O$_{4}$ (CdCr$_{2}$O$_{4}$) were found to be $\sim$8.125 ($\sim$8.079), $\sim$8.157 ($\sim$8.111) and $\sim$8.426 ($\sim$8.386) {\AA}, respectively. The calculated values of cubic to tetragonal distortion (c/a) for ZnV$_{2}$O$_{4}$, MgV$_{2}$O$_{4}$ and CdV$_{2}$O$_{4}$ were found to be $\sim$0.996, $\sim$0.995 and $\sim$0.997, respectively. Similarly, its values were found to be $\sim$0.997, $\sim$0.994 and $\sim$0.998 for ZnCr$_{2}$O$_{4}$, MgCr$_{2}$O$_{4}$ and CdCr$_{2}$O$_{4}$. Such a distortion, with c/a$<$1 for ZnV$_{2}$O$_{4}$, MgV$_{2}$O$_{4}$, ZnCr$_{2}$O$_{4}$ and MgCr$_{2}$O$_{4}$ compounds were close to the experimental results. The present study clearly shows the importance of ionic sizes in inducing the cubic to tetragonal distortion in these spinels. The ambiguities between calculated and experimentally reported values of c/a for both CdV$_{2}$O$_{4}$ and CdCr$_{2}$O$_{4}$ are expected to improve by considering the above mechanisms. These mechanisms may be also responsible for deciding the other physical properties of these compounds.

\acknowledgments {S.L. is thankful to UGC, India, for financial support.}

\pagebreak

\section{Tables}

\begin{table}[ht]
\caption{The muffin-tin sphere radii (in Bohr) used in the calculations for AV$_{2}$O$_{4}$ and ACr$_{2}$O$_{4}$ (A=Zn, Mg and Cd) compounds.}
\centering 
\begin{tabular}{p{3cm} p{2.5cm} p{2.5cm} p{2.5cm} p{2.5cm} p{2.5cm} p{2.5cm}}
\hline
\hline
{Compound}&{Zn}&{Mg}&{Cd}&{V}&{Cr}&{O}\\[0.5ex]
\hline
ZnV$_2$O$_4$&1.85&-&-&1.95&-&1.60\\
MgV$_2$O$_4$&-&1.75&-&1.90&-&1.60\\
CdV$_2$O$_4$&-&-&2.1&1.95&-&1.60\\
ZnCr$_2$O$_4$&1.80&-&-&-&1.80&1.60\\
MgCr$_2$O$_4$&-&1.80&-&-&1.95&1.60\\
CdCr$_2$O$_4$&-&-&2.1&-&1.95&1.60\\ [1ex]
\hline
\hline
\end{tabular}
\label{table:the exp}
\end{table}

\begin{table}[ht]
\caption{Calculated equilibrium lattice parameter $a$ ({\AA}), percentage difference (\%) between calculated and experimental lattice parameter, tetragonal distortion (c/a), bulk modulus (GPa) and atomic coordinates for AV$_{2}$O$_{4}$ and ACr$_{2}$O$_{4}$ (A=Zn, Mg and Cd) compounds, where the experimentally observed values are shown in the bracket.}
\centering 
\begin{tabular}{p{2.2cm}p{3.0cm}p{1.8cm}p{2.49cm}p{1.5cm}p{5.5cm}}
\hline
\hline
{Compound}&{Lattice parameter {\it a} ({\AA})}&{Percentage difference (\%)}&{Tetragonal distortion (c/a)}&{Bulk modulus (GPa)}&{Atomic coordinates}\\[0.0ex]
\hline
ZnV$_2$O$_4$\cite{Reehuis}&8.125(8.4028)&3.3&0.996(0.9948)&289&O: 32e(x,x,x) x=0.257(0.2604) \\
MgV$_2$O$_4$\cite{Wheeler}&8.157(8.42022)&3.1&0.995(0.9941)&244&O1: 16e(x,x,x) x=0.387(0.38623) O2: 16e(x,x,x) x=0.871(0.86623) V: 16e(x,x,x) x=0.635(0.6251)\\
CdV$_2$O$_4$\cite{Onoda}&8.426(8.691)&3.0&0.997(0.9877)&230&O: 32e(x,x,x) x=0.267(0.2672)\\
ZnCr$_2$O$_4$\cite{Dutton,Kagomiya}&8.079(8.320721)&2.9&0.997(0.9981)&254&O: 32e(x,x,x) x=0.258(0.26157)\\
MgCr$_2$O$_4$\cite{Martin}&8.111(8.3329)&2.6&0.994(0.9979)&243&O: 32e(x,x,x) x=0.259(0.2612)\\
CdCr$_2$O$_4$\cite{Chung}&8.386(8.59093)&2.4&0.998(1.004)&233&O: 32e(x,x,x) x=0.269(0.2681)\\ [1ex]
\hline
\hline
\end{tabular}
\label{table:the exp}
\end{table}

\pagebreak

\begin{figure}
\caption{Total energy difference between the volume dependent energies and energy corresponding to the equilibrium volume [$\Delta$$E$=$E$(V)-$E$(V$_{\rm eq}$)] per formula unit versus unit cell volume plots for (a) ZnV$_{2}$O$_{4}$, (b) MgV$_{2}$O$_{4}$, (c) CdV$_{2}$O$_{4}$, (d) ZnCr$_{2}$O$_{4}$, (e) MgCr$_{2}$O$_{4}$ and (f) CdCr$_{2}$O$_{4}$ compounds.}
\includegraphics{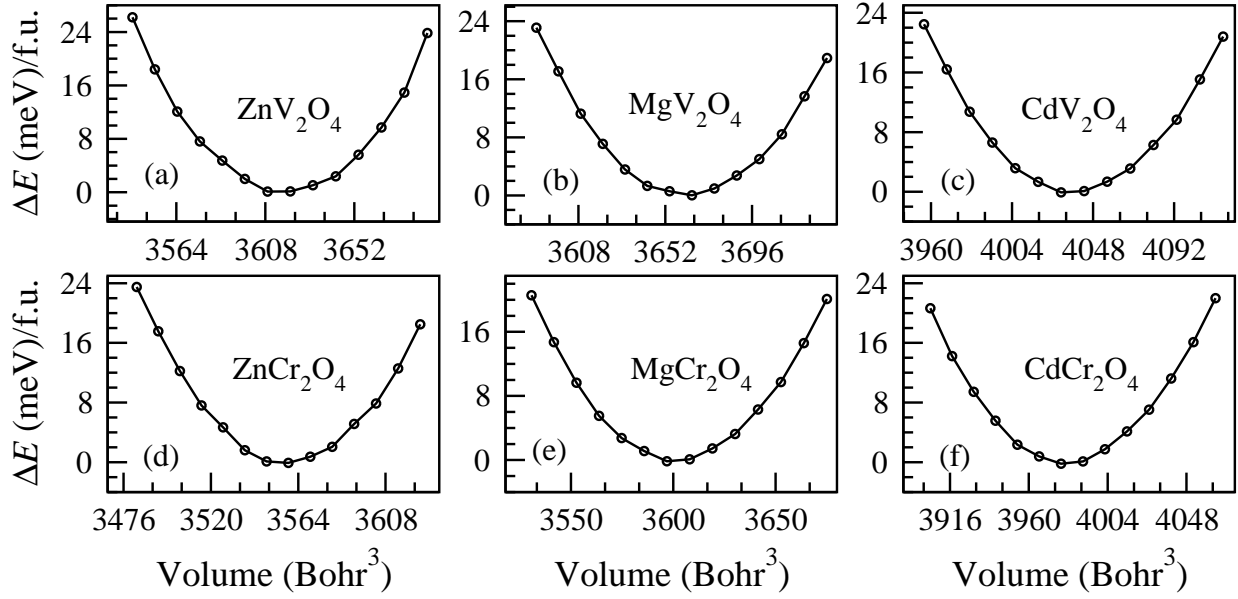}
\end{figure}

\begin{figure}
\caption{Total energy difference between the volume dependent energies and energy corresponding to the equilibrium volume [$\Delta$$E$=$E$(V)-$E$(V$_{\rm eq}$)] per formula unit versus percentage change in tetragonal distortion (c/a\%) plots for (a) ZnV$_{2}$O$_{4}$, (b) MgV$_{2}$O$_{4}$, (c) CdV$_{2}$O$_{4}$, (d) ZnCr$_{2}$O$_{4}$, (e) MgCr$_{2}$O$_{4}$ and (f) CdCr$_{2}$O$_{4}$ compounds.}
\includegraphics{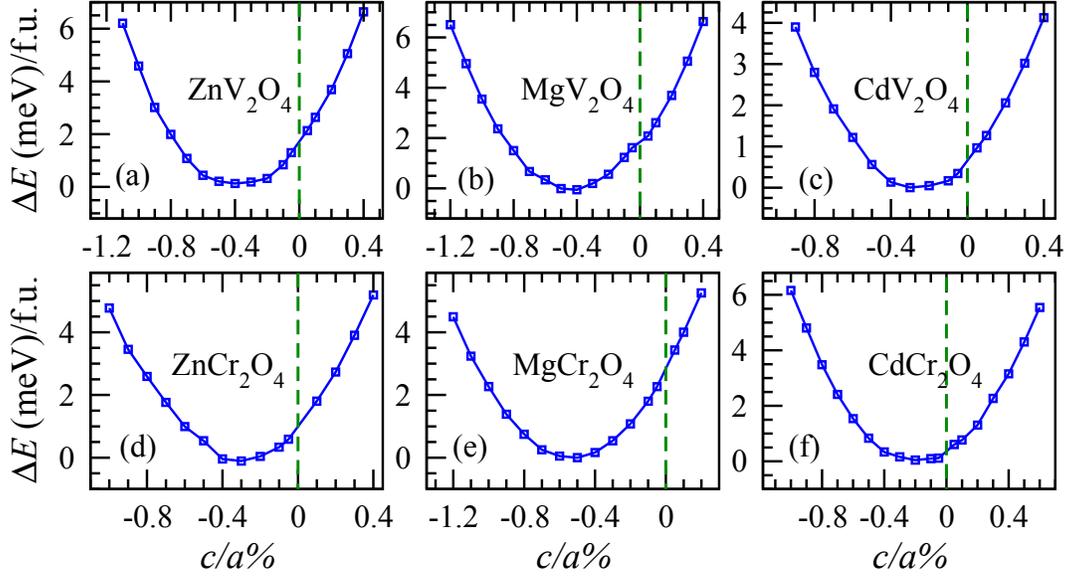}
\end{figure}

\begin{figure}
\caption{Partial density of states plots of V atom obtained in spin unpolarized calculations of both cubic and tetragonal structures (corresponding to equilibrium values of lattice parameters and atomic coordinates) of (a) ZnV$_{2}$O$_{4}$, (b) MgV$_{2}$O$_{4}$ and (c) CdV$_{2}$O$_{4}$ compounds. Zero energy corresponds to the Fermi level.} 
\includegraphics{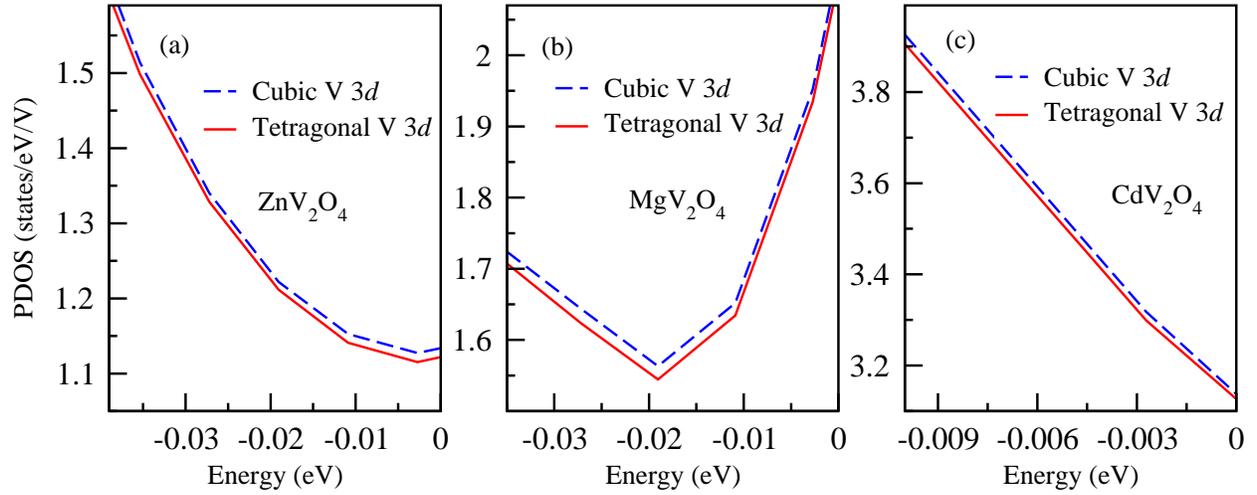}
\end{figure}

\begin{figure}
\caption{Partial density of states plots of Cr atom obtained in spin unpolarized calculations of both cubic and tetragonal structures (corresponding to equilibrium values of lattice parameters and atomic coordinates) of (a) ZnCr$_{2}$O$_{4}$, (b) MgCr$_{2}$O$_{4}$ and (c) CdCr$_{2}$O$_{4}$ compounds. Zero energy corresponds to the Fermi level.}
\includegraphics{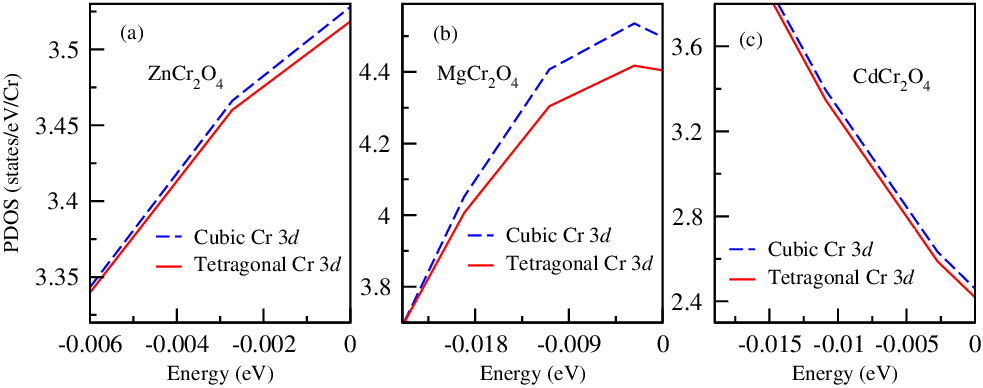}
\end{figure}


\begin{thebibliography}{99}

\bibitem{Goodenough} J. B. Goodenough, {\it Magnetism and the Chemical Bond}, Vol. I, (Interscience, New York, 1963).

\bibitem{Khomskii2014} D. I. Khomskii , {\it Transition Metal Compounds}, (Cambridge University Press, United Kingdom, 2014).

\bibitem{Kugel} K. I. Kugel$\acute{}$, and D. I. Khomski$\acute{\rm \oldstylenums{1}}$, Usp. Fiz. Nauk {\bf 136}, 621 (1982) [Sov. Phys. Usp. {\bf 25}, 231 (1982)].

\bibitem{Dutton} S. E. Dutton, Q. Huang, O. Tchernyshyov, C. L. Broholm, and R. J. Cava, Phys. Rev. B {\bf 83}, 064407 (2011), and references therein.

\bibitem{Mamiya} H. Mamiya, M. Onoda, T. Furubayashi, J. Tang, and I. Nakatani, J. Appl. Phys. {\bf 81}, 5289 (1997).

\bibitem{Nishiguchi} N. Nishiguchi, and M. Onoda, J. Phys: Condens. Matter {\bf 14}, L551 (2002).

\bibitem{Onoda}  M. Onoda, and J. Hasegawa, J. Phys.: Condens. Matter {\bf 15}, L95 (2003).

\bibitem{Reehuis} M. Reehuis, A. Krimmel, N. B$\ddot{\rm u}$ttgen, A. Loidl, and A. Prokofiev, Eur. Phys. J. B {\bf 35}, 311 (2003).

\bibitem{Yamashita} Y. Yamashita, and K. Ueda, Phys. Rev. Lett. {\bf 85}, 4960 (2000).

\bibitem{Tsunetsugu} H. Tsunetsugu, and Y. Motome, Phys. Rev. B {\bf 68}, 060405(R) (2003).

\bibitem{Lee} S. -H. Lee, D. Louca, H. Ueda, S. Park, T. J. Sato, M. Isobe, Y. Ueda, S. Rosenkranz, P. Zschack, J. $\acute{\rm I}$$\tilde{\rm n}$iguez, Y. Qiu, and  R. Osborn, Phys. Rev. Lett. {\bf 93}, 156407 (2004).

\bibitem{Tchernyshyov} O. Tchernyshyov, Phys. Rev. Lett. {\bf 93}, 157206 (2004).

\bibitem{Matteo} S. Di Matteo, G. Jackeli, and N. B. Perkins, Phys. Rev. B {\bf 72}, 020408(R) (2005).

\bibitem{Khomskii} D. I. Khomskii, and T. Mizokawa, Phys. Rev. Lett. {\bf 94}, 156402 (2005).

\bibitem{Suzuki} T. Suzuki, M. Katsumura, K. Taniguchi, T. Arima, and T. Katsufuji, Phys. Rev. Lett. {\bf 98}, 127203 (2007). 

\bibitem{Maitra} T. Maitra, and R. Valent$\acute{\rm \oldstylenums{1}}$, Phys. Rev. Lett. {\bf 99}, 126401 (2007).

\bibitem{Wheeler} E. M. Wheeler, B. Lake, A. T. M. N. Islam, M. Reehuis, P. Steffens, T. Guidi, and A. H. Hill, Phys. Rev. B {\bf 82}, 140406(R) (2010).

\bibitem{Giovannetti} G. Giovannetti, A. Stroppa, S. Picozzi, D. Baldomir, V. Pardo, S. Blanco-Canosa, F. Rivadulla, S. Jodlauk, D. Niermann, J. Rohrkamp, T. Lorenz, S. Streltsov, D. I. Khomskii, and J. Hemberger, Phys. Rev. B {\bf 83}, 060402(R) (2011). 

\bibitem{Pandey2011} S. K. Pandey, Phys. Rev. B {\bf 84}, 094407 (2011).

\bibitem{Pandey2012} S. K. Pandey, Phys. Rev. B {\bf 86}, 085103 (2012).

\bibitem{Lal2014} S. Lal, and S. K. Pandey, Eur. Phys. J. B {\bf 87}, 197 (2014).

\bibitem{Lal2016} S. Lal, and S. K. Pandey, J. Magn. Magn. Mater. {\bf 412}, 23 (2016).

\bibitem{Shaked} H. Shaked, J. M. Hastings, and L. M. Corliss, Phys. Rev. B {\bf 1}, 3116 (1970).

\bibitem{Broholm1} S. -H. Lee, C. Broholm, T. H. Kim, W. RatcliffII, and S. -W. Cheong, Phys. Rev. Lett. {\bf 84}, 3718 (2000).

\bibitem{Broholm2} S. -H. Lee, C. Broholm, W. Ratcliff, G. Gasparovic, Q. Huang, T. H. Kim, and S. -W. Cheong, Nature (London) {\bf 418}, 247204 (2002).

\bibitem{Rovers} M. T. Rovers, P. P. Kyriakou, H. A. Dabkowska, G. M. Luke, M. I. Larkin, and A. T. Savici, Phys. Rev. B {\bf 66}, 174434 (2002).

\bibitem{Moessner} O. Tchernyshyov, R. Moessner, and S. L. Sondhi, Phys. Rev. Lett. {\bf 88}, 067203 (2002).

\bibitem{Kagomiya} I. Kagomiya, H. Sawa, K. Siratori, K. Kohn, M. Toki, Y. Hata, and E. Kita, Ferroelectrics {\bf 268}, 327 (2002).

\bibitem{Martin} L. Ortega-San-Mart$\acute{\rm \oldstylenums{1}}$n, A. J. Williams, C. D. Gordon, S. Klemme, and J. P. Attfield, J. Phys.: Condens. Matter {\bf 20}, 104238 (2008).

\bibitem{Takagi} H. Takagi and S. Niitaka, {\it Introduction to Frustrated Magnetism}, Springer Series in Solid-State Sciences, Vol. 164, (Springer, Berlin, 2011), Part 3, p. 155.

\bibitem{Chung} J. -H. Chung, and Y. S. Song, J. Korean Phys. Soc. {\bf 62}, 12 (2013).

\bibitem{Ashcroft} N. W. Ashcroft, and N. D. Mermin, {\it Solid State Physics}, (Cengage Learning, New Delhi, 2010).

\bibitem{elk} http://elk.sourceforge.net.

\bibitem{Perdew} J. P. Perdew, and Y. Wang, Phys. Rev. B. {\bf 45}, 13244 (1992).

\bibitem{Vinett} P. Vinett, J. H. Rose, J. Ferrante, and J .R. Smith, J. Phys.: Condens. Matter {\bf 1}, 1941 (1989).

\bibitem{Haas} P. Haas, F. Tran, and P. Blaha, Phys. Rev. B {\bf 79}, 085104 (2009). 

\bibitem{Canosa} S. B. Canosa, F. Rivadulla, V. Pardo, D. Baldomir, J. -S. Zhou, M. G -Hern$\acute{\rm \oldstylenums{\rm a}}$ndez, M. A. L. -Quintela, J. Rivas, and J. B. Goodenough, Phys. Rev. Lett. {\bf 99}, 187201 (2007).

\bibitem{Catti} M. Catti, F. F. Fava, C. Zicovich, and R. Dovesi, Phys. Chem. Minerals {\bf 26}, 389 (1999).

\bibitem{Yong} W. Yong, S. Botis, S. R. Shieh, W. Shi, and A. C. Withers, Phys. Earth Planet. Inter. {\bf 196}, 75 (2012).  

\bibitem{Csonka} G. I. Csonka, J. P. Perdew, A. Ruzsinszky, P. H. T. Philipsen, S. Leb$\grave{\rm e}$gue, J. Paier, O. A. Vydrov, and J. G. $\acute{\rm A}$ngy$\acute{\rm a}$n, Phys. Rev. B {\bf 79}, 155107 (2009).

\bibitem{Lejaeghere} K. Lejaeghere $et$ $al$., Science {\bf 351}, aad3000 (2016).

\bibitem{Reichmann} H. J. Reichmann, and S. D. Jacobsen, Am. Miner. {\bf 89}, 1061 (2004). 

\bibitem{Haavik} C. Haavik, S. St$\varnothing$len, H. Fjellv$\mathring{\rm a}$g, M. Hanfland, and D. H$\ddot{\rm a}$usermann, Am. Miner. {\bf 85}, 514 (2000). 

\end{thebibliography}
\end{document}